# Overview of EIREX 2010: Computing

Julián Urbano, Mónica Marrero, Diego Martín and Jorge Morato

University Carlos III of Madrid
Department of Computer Science
Avda. Universidad, 30
Leganés, Madrid, Spain

## 1. Introduction

The first Information Retrieval Education through Experimentation track (EIREX 2010) was run at the University Carlos III of Madrid, during the 2010 spring semester.

EIREX 2010 is the first in a series of experiments designed to foster new Information Retrieval (IR) education methodologies and resources, with the specific goal of teaching undergraduate IR courses from an experimental perspective. For an introduction to the motivation behind the EIREX experiments, see the first sections of [Urbano et al., 2011]. For information on other editions of EIREX and related data, see the website at http://ir.kr.inf.uc3m.es/eirex/.

The EIREX series have the following goals:

- To help students get a view of the Information Retrieval process as they would find it in a real-world scenario, either industrial or academic.
- To make students realize the importance of laboratory experiments in Computer Science and have them initiated in their execution and analysis.
- To create a public repository of resources to teach Information Retrieval courses.
- To seek the collaboration and active participation of other Universities in this endeavor.

This overview paper summarizes the results of the EIREX 2010 track, focusing on the creation of the test collection and the analysis to assess its reliability. Next section provides a brief overview of our course and the student systems. Section 3 describes the process we followed to create the EIREX 2010 test collection, and Section 4 presents the evaluation results. Sections 5 and 6 analyze the reliability of our approach by studying the effects of the inconsistency and incompleteness of judgments. Section 7 wraps up with the conclusions.

## 2. Teaching Methodology

EIREX 2010 took place during the 2010 spring semester, in the context of the Information Retrieval and Access course [Urbano et al., 2010b], which is an elective course taken by senior Computer Science undergraduates. In this course we teach traditional IR techniques, and the main lab assignment consists in the development, from scratch, of a search engine for HTML documents. The development of this search engine is divided in three modules to hand in separately:

- Module 1 contains the implementation of an indexer for a collection of HTML documents and a simple retrieval model for automatic ad hoc queries.
- Module 2 incorporates query expansion in the retrieval process.
- Module 3 adds simple Named Entity Recognition (NER) capabilities to aid in the "who" questions.

In this edition we had 32 students, who created a total of 8 systems in groups of 4 students each. Thus, we had a total of 24 systems: 8 with basic retrieval, 8 with query expansion and 8 with NER. We try to encourage students by giving one extra point to the group who developed the most effective search engine (see Section 4).

All these systems are evaluated with an IR test collection built with the students also from scratch (see Section 3). A test collection for Information Retrieval evaluation contains three major components: a document collection, a set of information needs (usually called topics), and the relevance judgments or ground truth (usually assessed by humans) telling what documents are relevant to the topics [Voorhees, 2002]. The students run their systems for each topic, returning the list of documents in the collection deemed relevant to it. Then, we use some effectiveness measures to assess, according to the relevance judgments, how well the systems actually answered the information needs. This provides us with a ranking of the student systems in terms of effectiveness.

Ideally, students would be given a training collection to develop and tune their systems, and then they would be evaluated with a different test collection. However, in this first EIREX edition we did not have a full collection available for training, and so a subset of the final test collection was provided as training data (see Section 3.1). Students were thus given a subcollection, which was used to develop their first module. We then evaluated them with the full collection and published the results. The same process was repeated for modules two and three.

## 3. Test Collection

The process we employ to create the EIREX test collections is different from those usually followed in other IR evaluation workshops such as the early ad hoc tracks of the Text REtrieval Conference (TREC) ran by NIST [Voorhees et al., 2005]. Although we follow very similar principles, working with students bears some limitations, mainly in terms of effort and reliability. The document collection cannot be as large as those usually employed in TREC, because undergraduate students do not have the adequate expertise to handle that much information and they would probably dedicate too much time to efficiency issues rather than effectiveness and the implementation and understanding of the IR techniques we explain. This limitation restricts the topics to use: if they had nothing in common, we would probably need too many documents to have sufficient diversity to include relevant material for each topic; but if they were somehow similar, probably fewer documents would be needed. Therefore, we decided that all topics should have a common theme, which in addition reflects more closely a real setting where students have to deploy an IR system for a company in a particular domain. For this first EIREX edition we chose the theme to be *Computing*, as it sure is a topic attractive to Computer Science students. Thus, the document collection depends on the topics and not the other way around as usual.

Based on the theme chosen, we come up with a set of candidate topics. The problem at this point is how to build a document collection making sure that some relevant material is included for every topic. The procedure chosen consists in issuing queries to Google Web Search just as if we were trying to satisfy the information needs ourselves, manually using term proximity operators, query expansion, etc. Based on these searches and a brief inspection of their first results, we can discard topics apparently too difficult, with very few documents, or for which there do not seem to be clearly relevant documents. Once the final topic set is established, we use a focused web crawler [Urbano et al., 2010a], to download all web pages returned by Google for each topic. The union of all these web pages conform the *complete* document collection.

At this point we have a document collection and a set of topics, so next we need relevance judgments. Another difference here is that students have to make all relevance judgments *before* they start developing, as otherwise some might try cheating and judge all documents retrieved by their system as relevant. In addition, having them inspect the documents to assess their relevance helps later on during development because they know what kind of documents their systems will have to handle. Judging every document for every topic is completely impractical because it requires too much effort, so instead a sample of documents is judged for each topic (i.e. the topics' pools). To come up with a reliable pool of documents despite student systems do not directly contribute, we use well-known and freely available IR tools instead: Lemur[1] and Lucene[2] (call these the *pooling* systems). We thus proceed to index the complete document collection and obtain the results provided by different configurations of the pooling systems for each of the topics, trying to exploit as much as possible our previous knowledge about the topic and the information documents must contain to be considered relevant. For instance, if the topic asked for information about the CEO of a company, we would include the name of the person in the query. With these results we come up with the pools of documents students have to.

Pools are formed differently too: if we calculate *depth-k* pools (joining the top *k* results from the pooling systems), some topics might have considerably more documents to judge than others, as the final number depends on the overlap among the results. If some students were assigned a pool significantly larger than others, they could just judge carelessly once they think they have done enough work compared to their classmates. To prevent this situation we compute *size-k* pools instead: pools with the minimum depth such that the total size is at least *k* documents. Thus, each topic has a pool of documents with different depth, although all pools have very similar sizes and so all students judge more or less the same amount of documents.

---

[1] http://www.lemurproject.org
[2] http://lucene.apache.org

| Topic | Downloaded | Pool size | Pool depth | Kappa | Overlap | Precision | Recall |
|---|---|---|---|---|---|---|---|
| 001 | 328 | 100 | 28 | 0.362 | 0.484 | 0.811 | 0.545 |
| 002 | 417 | 100 | 26 | 0.243 | 0.233 | 0.233 | 1.000 |
| 003 | 616 | 100 | 30 | 0.517 | 0.763 | 0.906 | 0.829 |
| 004 | 220 | 102 | 25 | 0.470 | 0.536 | 0.769 | 0.638 |
| 005* | 417 | 101 | 17 | - | - | - | - |
| 006** | 768 | 102 | 19 | 0.468 | 0.548 | 0.622 | 0.821 |
| 007 | 547 | 100 | 25 | 0.456 | 0.407 | 0.889 | 0.429 |
| 008 | 729 | 100 | 23 | 0.096 | 0.158 | 0.818 | 0.164 |
| 009 | 374 | 100 | 37 | 0.625 | 0.550 | 1.000 | 0.550 |
| 010 | 609 | 101 | 26 | 0.217 | 0.111 | 1.000 | 0.111 |
| 011 | 218 | 100 | 56 | 0.192 | 0.200 | 0.250 | 0.500 |
| 012* | 338 | 100 | 19 | - | - | - | - |
| 013 | 384 | 100 | 21 | 0.333 | 0.269 | 0.269 | 1.000 |
| 014 | 247 | 100 | 58 | 0.342 | 0.403 | 0.595 | 0.556 |
| 015 | 435 | 102 | 34 | 0.624 | 0.667 | 0.810 | 0.791 |
| 016 | 417 | 103 | 28 | 0.433 | 0.444 | 0.526 | 0.741 |
| 017 | 516 | 101 | 23 | 0.574 | 0.364 | 0.500 | 0.571 |
| 018 | 474 | 101 | 20 | 0.735 | 0.708 | 0.895 | 0.773 |
| 019* | 488 | 100 | 15 | - | - | - | - |
| 020 | 459 | 105 | 20 | 0.395 | 0.263 | 0.278 | 0.833 |
| 021† | 79 | - | - | - | - | - | - |
| 022† | 689 | - | - | - | - | - | - |
| Average | 444 | 101 | 28 | 0.417 | 0.418 | 0.649 ± 0.054 | 0.646 ± 0.054 |
| Total | 9,769 | 1,967 | - | - | - | - | - |

Table 1. Summary of the EIREX 2010 test collection. * for topics judged by one faculty member, ** judged by two faculty members. † for the noise topics.

Although unlikely, the results provided by the pooling systems might still leave out relevant documents. To assure that all pools have some relevant material, we always include Google's top $k_G$ results for each topic, as we checked when selecting topics that some relevant web pages were included there. Also, we add $k_N$ random documents crawled from noise topics, which we created by excluding specific terms appearing in the topic set descriptions. These noise documents allow us to check for quality in the relevance judgments, as they should all be judged not relevant for any topic. If we found students judging these noise documents as relevant, we would have an indication of possible negligence. Therefore, all pools have $k_N$ noise documents, the first $k_G$ documents retrieved by Google, and documents retrieved by the pooling systems up to a minimum of $k$ documents altogether. The union of all documents in these pools conform the *biased* document collection. This is the collection we provide students with to run and evaluate their systems.

### 3.1. Topics

The EIREX 2010 test collection contains a total of 20 topics, all of which pertain to the *Computing* theme we chose. All topic descriptions have a common structure (see Figure 1), with a unique id, a title and a description of what is considered to be relevant to the topic. To keep things simple in this first edition, we decided to have all topics share the same generic description of relevance levels (see Section 3.4).

```
<topic id="2010-019">
  <title>Where are Google's data-centers located?</title>
  <relevance>
    <level value="2">The document is not related to the topic. It may contain some common terms, but still not related
        to the topic.</level>
    <level value="1">The document is related to the topic, but does not satisfy the information need. It may contain a
        hyperlink to a relevant document.</level>
    <level value="0">The document is related to the topic and does satisfy the information need.</level>
  </relevance>
</topic>
```

Figure 1. A sample EIREX 2010 topic description.

The topic titles were used as input queries to the student systems, so they can all be considered *short automatic* runs in TREC's terminology [Voorhees et al., 2005] (i.e. there is no human intervention in creating the queries from the topic descriptions). They are rather short, with only 9 words on average, ranging from 5 to 14 words. Due to the lack of a training collection, this year we provided students with a very small subset containing topics 005, 006, 012 and 019, for which we made the relevance judgments ourselves (see Section 3.4). Topics 021 and 022 were used as noise topics to obtain nonrelevant documents[3].

### 3.2. Documents

The *complete* document collection contains all documents returned by Google for the final set of 20 topics plus the 2 noise topics (see Table 1). A total of 9,769 web pages were crawled for all 22 topics, which account for 735 MB. The median size per document is 49 KB, with a mean of 77 KB. The median number of words per document is 1,307, with a mean of 2,668. These documents were used just as downloaded, with no postprocessing involved.

The *biased* collection, containing only documents in the pools (see Section 3.3), had a total of 1,967 documents, which account for 161 MB. The median size per document is 44 KB, with a mean of 84 KB. The median number of words per document is 1,319, with a mean of 3,119. That is, both document collections have roughly the same characteristics in terms of size and word count, though the biased collection is a little more skewed.

### 3.3. Pools

For each of the 20 topics in the collection, we ran the 12 pooling systems described in Table 2. We used various configurations of Lemur version 4.11 and Lucene version 3.0.1, which basically differed on the stemmer, the treatment of stop words, the retrieval model employed and the use of query expansion.

| Id | System | Parse HTML | Stemmer | Stop words | Model | Query expansion |
|---|---|---|---|---|---|---|
| p0001 | Lemur 4.11 | Yes | Krovetz | No | Okapi BM25 | No |
| p0002 | Lemur 4.11 | Yes | Krovetz | No | Okapi BM25 | Yes |
| p0003 | Lemur 4.11 | Yes | Krovetz | Yes | Okapi BM25 | No |
| p0004 | Lemur 4.11 | Yes | Krovetz | Yes | Okapi BM25 | Yes |
| p0005 | Lemur 4.11 | Yes | No | No | Okapi BM25 | No |
| p0006 | Lemur 4.11 | Yes | No | No | Okapi BM25 | Yes |
| p0007 | Lemur 4.11 | Yes | No | Yes | Okapi BM25 | No |
| p0008 | Lemur 4.11 | Yes | No | Yes | Okapi BM25 | Yes |
| p0009 | Lucene 3.0.1 | No | No | Yes | Vectorial TF/IDF | No |
| p0010 | Lucene 3.0.1 | Yes | No | Yes | Vectorial TF/IDF | Yes |
| p0011 | Lucene 3.0.1 | Yes | Porter | Yes | Vectorial TF/IDF | No |
| p0012 | Lucene 3.0.1 | Yes | Porter | Yes | Vectorial TF/IDF | Yes |

Table 2. Summary of the EIREX 2010 pooling systems.

For each topic, we joined the top $k_G$ documents retrieved by Google and $k_N$ random documents from the two noise topics. In this EIREX 2010 edition we chose $k_G=k_N=10$ documents. Then, we pooled results from the 12 pooling systems until at least 100 documents were in the pool altogether. As shown in Table 1, pool sizes ranged between 100 and 105, with an average of 101 documents. Therefore, all students judged more or less the same amount of documents. Pool depths ranged between 15 and 58, with an average 28, showing that the pooling systems tended to agree much more for some topics than for others. Note that the biased collection contains 1,967 unique documents, but the sum of all pool sizes is 2,018. Therefore, several documents were retrieved for more than one topic: 49 were retrieved for 2 topics and 1 was retrieved for 3 topics.

### 3.4. Relevance Judgments

We applied a cleaning process to all web pages before being displayed to the assessors, turning them into a basic black and white document to make the reading task easier. We also removed all scripts, embedded objects and HTML elements not related with page rendering. Assessors were able to use a basic search option, and they of course did not know whether documents were from the Google top results or noise topics. Judging took about 2 hours per assessor and topic, so the task could be completed in about one class session. Students never had access to the relevance scores of the documents, as all files were encrypted for submission back to the course instructors.

---

[3] Noise topics have no description in the topics file.

All ~100 documents per topic were judged by two students, except for topics 005, 012 and 019, which were judged by one faculty member; and topic 006, which was judged by two faculty members. We used a 3-point relevance scale from 0 to 2: nonrelevant, somewhat relevant, and highly relevant. Documents that could not be judged due to technical problems when rendering were judged as -1 (1% of the times). On average, students judged 24 documents per topic as somewhat relevant and 13 as highly relevant. In addition, of all the 326 judgments on noise documents, only once did a student judge the document as relevant.

## 4. Evaluation Results

All 20 topics were used to evaluate the 24 student systems (3 modules for each of the 8 student groups). We used NDCG@100 (Normalized Discounted Cumulated Gain at 100 documents retrieved) as the main measure to rank systems, and AP@100 (Average Precision), P@10 (Precision) and RR (Reciprocal Rank) as secondary measures, using a 2-point relevance scale conflating the somewhat and highly relevant levels.

| System | NDCG@100 | AP@100 | P@10 | RR |
|---|---|---|---|---|
| 03.3 | 0.699 ± 0.022 | 0.536 ± 0.029 | 0.621 ± 0.039 | 0.745 ± 0.053 |
| 03.2 | 0.685 ± 0.021 | 0.52 ± 0.029 | 0.598 ± 0.038 | 0.712 ± 0.049 |
| 03.1 | 0.683 ± 0.022 | 0.519 ± 0.029 | 0.583 ± 0.039 | 0.711 ± 0.037 |
| 01.1 | 0.683 ± 0.022 | 0.51 ± 0.026 | 0.553 ± 0.030 | 0.691 ± 0.038 |
| 01.3 | 0.68 ± 0.022 | 0.503 ± 0.026 | 0.538 ± 0.041 | 0.678 ± 0.037 |
| 01.2 | 0.671 ± 0.020 | 0.484 ± 0.025 | 0.53 ± 0.034 | 0.675 ± 0.048 |
| 05.3 | 0.661 ± 0.024 | 0.48 ± 0.027 | 0.53 ± 0.040 | 0.667 ± 0.049 |
| 07.1 | 0.66 ± 0.021 | 0.476 ± 0.032 | 0.528 ± 0.035 | 0.661 ± 0.045 |
| 05.1 | 0.653 ± 0.024 | 0.472 ± 0.031 | 0.528 ± 0.035 | 0.661 ± 0.058 |
| 07.3 | 0.652 ± 0.024 | 0.464 ± 0.031 | 0.528 ± 0.040 | 0.65 ± 0.049 |
| 02.1 | 0.634 ± 0.025 | 0.455 ± 0.029 | 0.517 ± 0.040 | 0.649 ± 0.055 |
| 05.2 | 0.629 ± 0.023 | 0.448 ± 0.030 | 0.513 ± 0.042 | 0.644 ± 0.052 |
| 07.2 | 0.628 ± 0.024 | 0.442 ± 0.031 | 0.505 ± 0.041 | 0.633 ± 0.054 |
| 04.1 | 0.528 ± 0.018 | 0.345 ± 0.029 | 0.42 ± 0.043 | 0.59 ± 0.041 |
| 08.1 | 0.522 ± 0.022 | 0.335 ± 0.019 | 0.381 ± 0.026 | 0.499 ± 0.038 |
| 08.2 | 0.508 ± 0.018 | 0.32 ± 0.018 | 0.366 ± 0.027 | 0.455 ± 0.046 |
| 02.3 | 0.508 ± 0.018 | 0.32 ± 0.018 | 0.36 ± 0.022 | 0.455 ± 0.046 |
| 02.2 | 0.507 ± 0.018 | 0.319 ± 0.018 | 0.345 ± 0.034 | 0.447 ± 0.020 |
| 08.3 | 0.486 ± 0.012 | 0.297 ± 0.013 | 0.345 ± 0.034 | 0.447 ± 0.020 |
| 04.2 | 0.472 ± 0.018 | 0.291 ± 0.023 | 0.321 ± 0.029 | 0.446 ± 0.039 |
| 04.3 | 0.472 ± 0.018 | 0.291 ± 0.023 | 0.321 ± 0.029 | 0.422 ± 0.045 |
| 06.3 | 0.349 ± 0.017 | 0.163 ± 0.015 | 0.295 ± 0.023 | 0.418 ± 0.045 |
| 06.1 | 0.347 ± 0.017 | 0.163 ± 0.015 | 0.295 ± 0.023 | 0.409 ± 0.044 |
| 06.2 | 0.325 ± 0.015 | 0.15 ± 0.012 | 0.282 ± 0.021 | 0.313 ± 0.016 |
| Mean σ | 0.020 | 0.024 | 0.034 | 0.043 |

Table 3. Mean and standard deviation of the NDCG@100, AP@100, P@10 and RR scores for the 24 student systems over 1,000 random combinations of trels, ordered by mean NDCG@100 score.

Given that most topics were judged by two different assessors, there is no single ground truth to evaluate systems; so we decided to create a sample by randomly combining assessors for each topic that was judged twice. There were 17 such topics, so there are $2^{17}$=131,072 possible combinations, though for practical reasons we created a sample of 1,000. Call each of these a *trel* (for topic relevance set). Table 3 and the plots in Figure 2 show the mean scores for each of the four measures over the 1,000 trels, along with the range of scores observed.

Systems behaved remarkably well compared to usual TREC ad hoc results [Voorhees et al., 2005], probably due to the methodology followed to build the test collection (see Section 3). Documents were crawled for a prefixed set of topics, and if topics were quite different from one another (which we attempted to avoid), documents would probably be very different too. That is, it might be somewhat clear, from an algorithmic perspective, what documents pertain to what topics, although systems would still have to rank relevant documents properly. This can be observed in Figure 3. The left plot shows, for each group's best system, the ratio of documents retrieved at different cutoffs that were crawled for the topic. Call this measure C@k (Crawl). We can see that for the top systems over 90% of the documents retrieved were actually crawled for the particular topics for *k*<60, and over 70% by the end cutoff *k*=100. The right plot displays the R@k scores (Recall), showing that the top systems retrieved about 90% of the relevant documents by the end cutoff $k$=100, indicating that systems did not have much trouble in finding the relevant material for each topic (we used the union trels here, see Section 5.2, as they contain the larger proportion of relevant judgments). Most importantly, we can see a direct relationship between the C@k and R@k scores and the ranking of systems, indicating that the top systems performed better because they retrieved more documents from Google's results and did not overlap with the documents crawled for other topics.

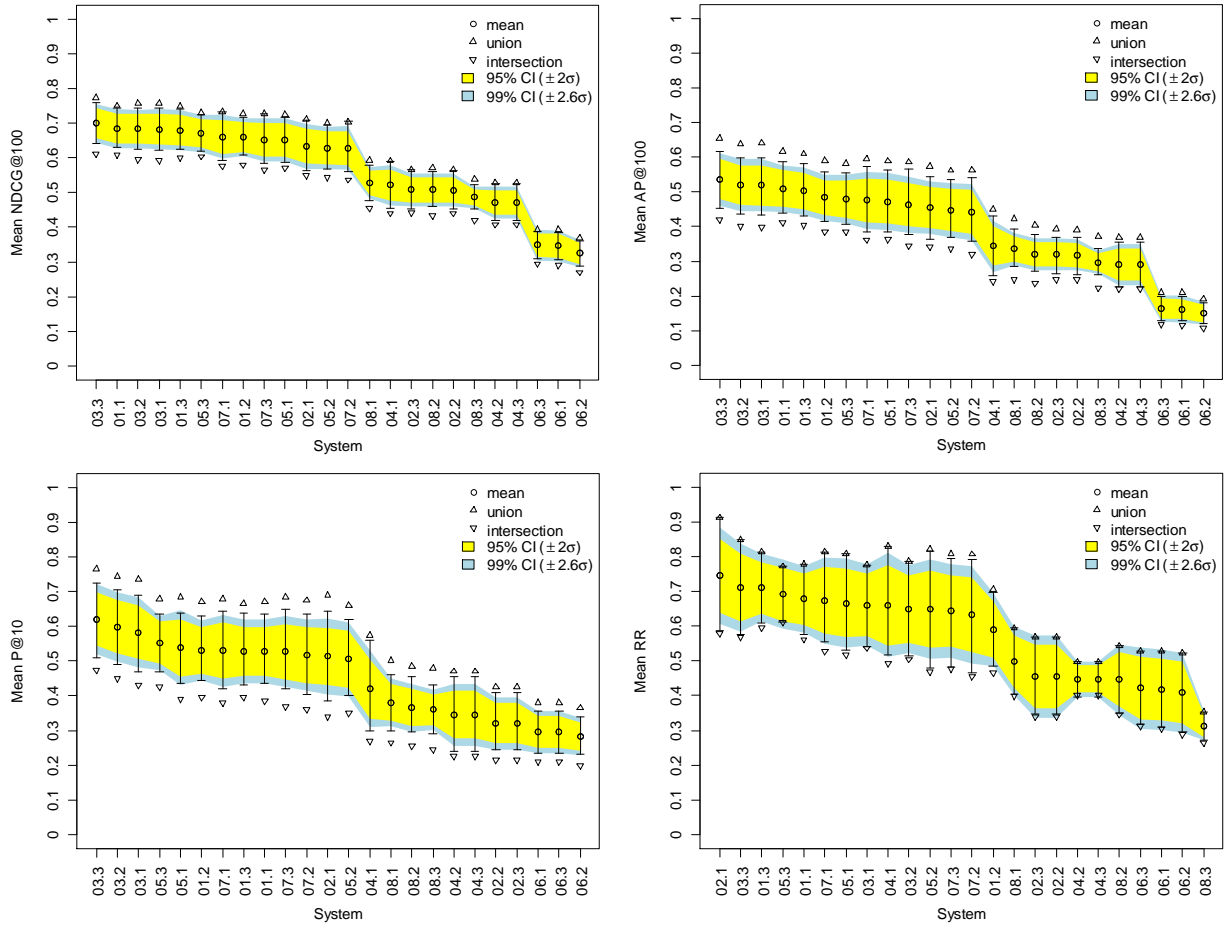

Figure 2. Mean NDCG@100 (top left), AP@100 (top right), P@10 (bottom left) and RR (bottom right) scores for the 24 student systems over 1,000 random combinations of trels, as well as the union and intersection trels. Errorbars show the range of scores observed, while the yellow and blue regions correspond to 2 and 2.6 standard deviations around the means (95% and 99% CI).

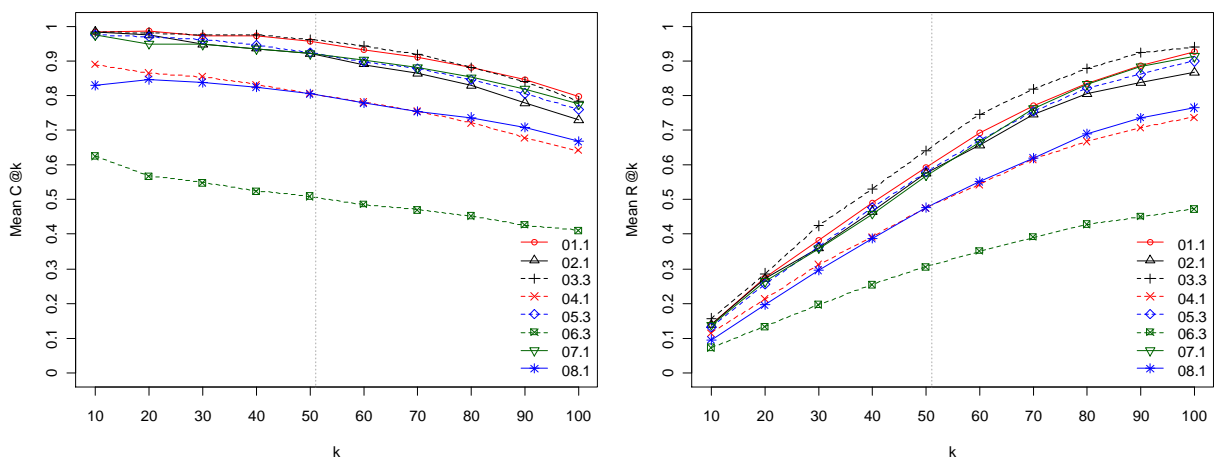

Figure 3. Mean C@k (left) and R@k (right) with the union trels for the best system per student group. The grey vertical line marks the mean number of relevant documents across topics (51).

# 5. Inconsistency of Relevance Judgments

One of the major critics of IR systems evaluations following TREC-like methodologies is the subjective nature of relevance, which can affect the results when using different assessors. It has been shown that such differences in relevance assessments do have an impact on the measured performance of systems, but that the ranking of systems is hardly altered, which is what matters given the comparative nature of these experiments [Voorhees, 2000][Voorhees, 2002]. We made a similar meta-analysis on our collection, trying to measure the reliability of the judgments made by students.

## 5.1. Assessor Agreement

In her study, Voorhees collected relevance judgments from 3 different people and for each of the 50 TREC-4 topics. She measured the average agreement between each topic's three assessors, calculating the overlap of judgments (the size of the intersection of relevant documents divided by the size of the union of relevant documents), and the precision and recall using one assessor's judgments as ground truth and the other's as a retrieval run (note that the precision of assessor A with respect to assessor B is the recall of B with respect to A and vice versa). She observed overlap levels between 0.301 and 0.494, precision ratios between 0.605 and 0.819 and recall ratios between 0.528 and 0.695 [Voorhees, 2000].

In our collection, we have 20 topics, 17 of which were evaluated by two different assessors. We also measured the agreement between each topic's assessors using overlap, precision and recall, besides Cohen's Kappa with equal weights (see Table 1). The average Kappa score is fairly high: 0.417 across all 17 topics. The mean overlap across topics is 0.418, which is highly within the range observed by Voorhees for TREC assessors. Finally, the precision and recall averages are 0.649 and 0.646 respectively, which agree with Voorhees' finding that a practical upper bound on performance is 65% precision at 65% recall, as that is the level at which humans tend to agree with one another [Voorhees, 2000]. However, these scores depend on who we consider assessor A and who assessor B, so we computed the precision-recall ratios over a random sample of 1,000 combinations. About 95% of the observations were between 0.539 and 0.757. Therefore, the judgments made by our students seem to be as reliable as those of TREC assessors, because the agreement scores are comparable.

## 5.2. Effect on System Performance

Using the 1,000 random trels from Section 4 we can measure the differences in effectiveness scores due to having different relevance assessors. We also created two special trels: the *union* (where a document has the largest relevance level given by either judge) and the *intersection* (with the lowest relevance level given by either judge). Note that the union trels model a very permissive assessor, while the intersection trels model a restrictive one. Therefore, we have 1,002 different trels to evaluate the 24 student systems. Figure 2 shows the results for mean NDCG@100, AP@100, P@10 and RR scores.

Table 4 shows the minimum and maximum largest differences observed across all systems when using various sets of trels: for each system we computed its minimum and maximum scores across all 1,000 trels, and then record the difference; the table reports the minimum and maximum such (largest) differences across all 24 systems. Differences in NDCG@100 are between 0.068 and 0.145, and between 0.061 and 0.189 for AP@100, which are much smaller than for P@10 and RR. These differences are larger than those observed by Voorhees (between 0.05 and 0.1 for mean AP in TREC-4). However, this is expected because in our case the relevance judgments were made in a 3-point scale, introducing more variability; the students have much less experience than TREC assessors; and we have less than half the number of topics used in TREC, resulting in more unstable results to begin with [Buckley et al., 2000]. Also, the mean AP@100 scores of the student systems are larger than those observed in TREC-4 in the first place (between 0 and 0.4), so the relative differences are virtually the same. As found by Voorhees, the changes in performance are highly correlated across topics: if a system gets a higher score with a particular set of trels, the other systems tend to do so.

|  | NDCG@100 | AP@100 | P@10 | RR |
| --- | --- | --- | --- | --- |
| 95% CI (± 2$\sigma$) | 0.048 - 0.1 | 0.05 - 0.128 | 0.082 - 0.170 | 0.064 - 0.231 |
| All 1,000 trels | 0.068 - 0.145 | 0.061 - 0.189 | 0.110 - 0.260 | 0.081 - 0.339 |
| union - intersection | 0.098 - 0.167 | 0.084 - 0.244 | 0.165 - 0.35 | 0.088 - 0.354 |

Table 4. Observed minimum and maximum largest system effectiveness differences when using various combinations of trels: around systems mean's 95% confidence intervals (top), between all 1,000 random trels (middle), and between the union and intersection trels (bottom).

However, most differences lie in a much narrower interval. The yellow shaded regions in Figure 2 represent the intervals within which 95% of the observations can be found per system (4 standard deviations long). The largest such interval is 0.1 for NDCG@100 and 0.128 for AP@100, which are quite small relative to the absolute effectiveness figures. The results in Table 3 (bottom row) and Table 4 support the use of NDCG@100 as the main measure to rank systems because it is shown to be the most stable under different assessors, followed by AP@100, P@10 and RR. These results agree with previous studies on the effect of topic set size on measure stability [Buckley et al., 2000][Sakai, 2007]. When comparing the union and intersection trels, differences are much larger as expected, especially for P@10 and RR, where there are observations over 0.35 (about half the average effectiveness).

### 5.3. Effect on System Ranking

We obtained 5,000 random pairs of trels and calculated the Kendall's τ correlation coefficient between the rankings of systems resulting from evaluating them with those two trels. Doing so, we measure how different the ranking of systems would be if using a different (yet possible) set of trels. As Table 5 shows, the correlations are quite high for NDCG@100 and AP@100, averaging to 0.926 and 0.927 respectively.

|          | NDCG@100       | AP@100         | P@10           | RR             |
|----------|----------------|----------------|----------------|----------------|
| Mean ± σ | 0.926 ± 0.028  | 0.927 ± 0.028  | 0.868 ± 0.048  | 0.739 ± 0.078  |
| Minimum  | 0.804          | 0.775          | 0.7            | 0.464          |
| Maximum  | 1              | 1              | 1              | 1              |

Table 5. Kendall's τ correlation between the system rankings resulting from 5,000 random pairs of trels.

In her study, Voorhees found correlations between 0.841 and 0.996 in mean AP, with an average of 0.938; concluding that correlations above 0.9 can be considered reliable [Voorhees, 2000]. In our case, the distribution of correlations for NDCG@100 and AP@100 are extremely similar, and over 80% of the correlations were larger than 0.9 for both measures. Indeed, none of the ranking swaps between two systems were significant (Wilconxon sign-rank test, $\alpha$=0.05). As expected, the correlations with P@10 and RR are much lower because there are more unstable measures.

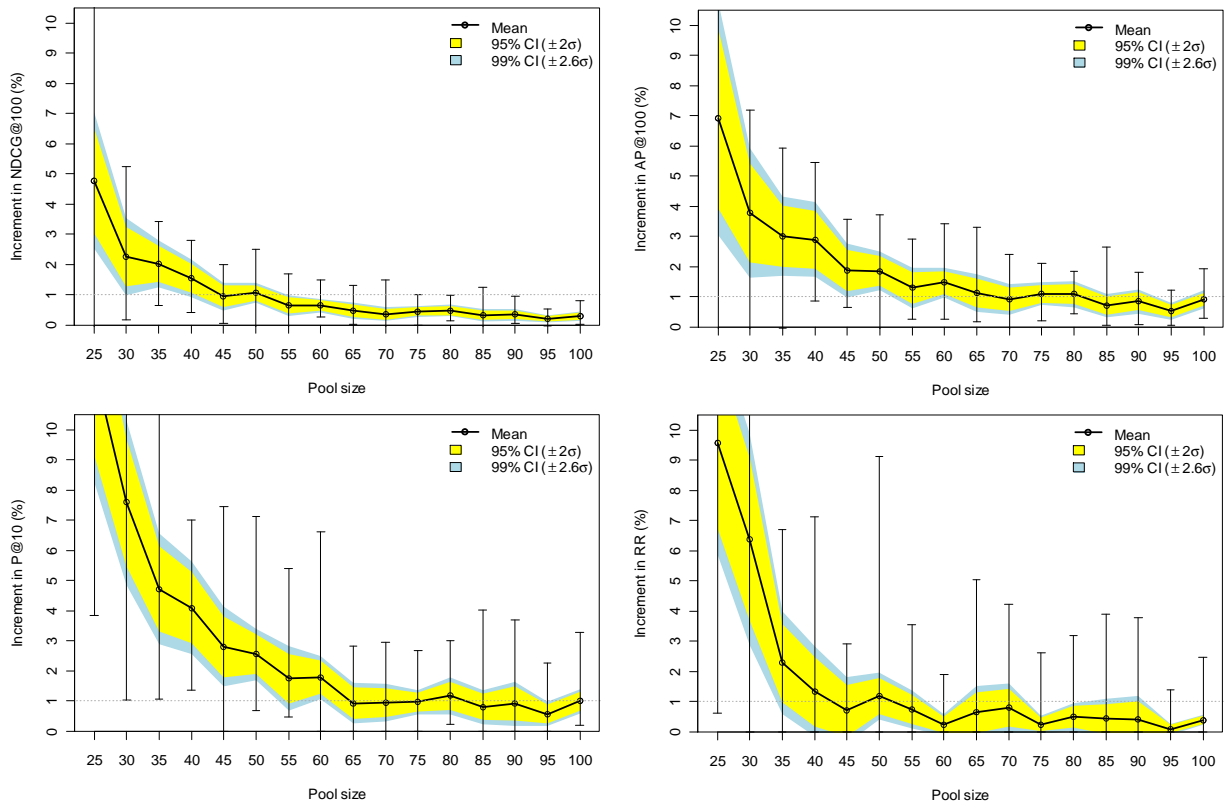

Figure 4. Mean NDCG@100 (top left), AP@100 (top right), P@10 (bottom left) and RR (bottom right) increments, over 1,000 random combinations of trels, as a function of pool size (lower is better). Error bars show the range of increments observed, while the yellow and blue regions correspond to 2 and 2.6 standard deviations around the means (95% and 99% CI).

# 6. Incompleteness of Relevance Judgments

Another drawback of TREC-like evaluations is that the sets of relevance judgments are incomplete, because only the documents in the pool are judged [Voorhees, 2002]. Of course, if a system does not have the opportunity to contribute much to the pool it is expected to have its effectiveness diminished, as it might have retrieved relevant material which is unknown. In the worst case, a brand new system using these collections did not contribute at all to the pools, and so the evaluation could be unreliable. This is our case, as the systems developed by the students did not contribute directly to the pools, only the Lemur and Lucene systems did (the pooling systems). The effect of incompleteness has been studied with TREC dada, concluding that the early ad hoc tracks were quite robust to the incompleteness problem [Zobel, 1998]. Next, we analyze our collection in the same line.

## 6.1. Effect on System Performance

We generated pools of size 20, with the 10 noise documents and the top 10 retrieved by Google for each topic. Then, we added documents from the pooling systems to a minimum pool size of 25, 30, 35, and so on, up to the final pools of at least 100 documents. This gives us 17 different pools, each of which can be evaluated with the 1,000 different trels from Section 4. We evaluated the 24 student systems for each pool and each trel. Then, for each increment of 5 documents in the pool, we calculated the difference in effectiveness for each system, between the two pools, and for all 1,000 trels. The difference is measured as the percentage increased in effectiveness from the smaller to the larger pool, so that it is directly comparable with Zobel's findings (differences between 0.5% and 3.5%, with some observations of up to 19% in TREC-3). Figure 4 shows how the average difference in effectiveness diminishes as the pool size increases.

In the case of NDCG@100, pool sizes larger than 55 show subsequent increments mostly below 1%, and no difference is observed over 2% (see Table 6). By the end, with pools of 100 documents, most increments are below 0.5% (the blue shaded area corresponds to a 99% confidence interval around the means). In the case of AP@100 the pool size needs to be at least 85 for the average difference to drop below 1%, while some differences above 2% can still be found. These results show that the pools seem to be reliable compared to TREC's, although a larger pool size would further increase the reliability of AP@100 and P@10. However, it is noticeable how well RR behaves on average. This is due to the fact that RR only needs one single relevant document (per topic) to compute the score, and those single documents are usually found at the top of the pools. Thus, the RR score differences rapidly converge towards zero. For other measures such as AP@100, more documents are needed for the score differences to converge, and so larger pools are required. Nonetheless, RR still shows the worst differences because of its stability issues: until the truly top relevant document retrieved is judged, scores with small pools can be very different to the real scores.

| Pool size | NDCG@100 Mean ± σ | Max | AP@100 Mean ± σ | Max | P@10 Mean ± σ | Max | RR Mean ± σ | Max |
|---|---|---|---|---|---|---|---|---|
| 20 → 25 | 4.77 ± 0.87 | 10.94 | 6.91 ± 1.49 | 13.54 | 12.07 ± 1.47 | 20.18 | 9.57 ± 1.42 | 18.34 |
| 25 → 30 | 2.25 ± 0.49 | 5.25 | 3.79 ± 0.82 | 7.19 | 7.6 ± 1.05 | 14.29 | 6.4 ± 1.35 | 14.99 |
| 30 → 35 | 2.02 ± 0.30 | 3.41 | 3 ± 0.50 | 5.93 | 4.72 ± 0.70 | 12.86 | 2.28 ± 0.65 | 6.72 |
| 35 → 40 | 1.54 ± 0.24 | 2.8 | 2.89 ± 0.48 | 5.46 | 4.09 ± 0.59 | 7.02 | 1.33 ± 0.57 | 7.12 |
| 40 → 45 | 0.93 ± 0.18 | 1.98 | 1.87 ± 0.34 | 3.57 | 2.8 ± 0.51 | 7.45 | 0.7 ± 0.42 | 2.92 |
| 45 → 50 | 1.05 ± 0.13 | 2.5 | 1.85 ± 0.25 | 3.71 | 2.55 ± 0.33 | 7.14 | 1.18 ± 0.30 | 9.14 |
| 50 → 55 | 0.64 ± 0.14 | 1.69 | 1.29 ± 0.26 | 2.92 | 1.74 ± 0.41 | 5.41 | 0.73 ± 0.24 | 3.56 |
| 55 → 60 | 0.63 ± 0.09 | 1.48 | 1.47 ± 0.19 | 3.43 | 1.79 ± 0.28 | 6.63 | 0.22 ± 0.14 | 1.91 |
| 60 → 65 | 0.46 ± 0.10 | 1.29 | 1.11 ± 0.24 | 3.3 | 0.92 ± 0.26 | 2.83 | 0.64 ± 0.34 | 5.05 |
| 65 → 70 | 0.36 ± 0.09 | 1.48 | 0.92 ± 0.20 | 2.42 | 0.94 ± 0.24 | 2.94 | 0.79 ± 0.32 | 4.24 |
| 70 → 75 | 0.44 ± 0.07 | 1.02 | 1.1 ± 0.14 | 2.12 | 0.97 ± 0.16 | 2.69 | 0.23 ± 0.11 | 2.61 |
| 75 → 80 | 0.47 ± 0.07 | 0.97 | 1.08 ± 0.16 | 1.84 | 1.17 ± 0.24 | 2.99 | 0.48 ± 0.18 | 3.17 |
| 80 → 85 | 0.31 ± 0.08 | 1.25 | 0.71 ± 0.15 | 2.64 | 0.8 ± 0.22 | 4.02 | 0.42 ± 0.25 | 3.89 |
| 85 → 90 | 0.34 ± 0.07 | 0.93 | 0.85 ± 0.15 | 1.82 | 0.91 ± 0.28 | 3.68 | 0.41 ± 0.30 | 3.77 |
| 90 → 95 | 0.2 ± 0.05 | 0.53 | 0.52 ± 0.11 | 1.22 | 0.57 ± 0.15 | 2.25 | 0.09 ± 0.07 | 1.39 |
| 95 → 100 | 0.28 ± 0.06 | 0.8 | 0.91 ± 0.11 | 1.92 | 0.99 ± 0.15 | 3.26 | 0.39 ± 0.07 | 2.47 |

Table 6. Mean, standard deviation and maximum of the percentage increments in NDCG@100, AP@100, P@10 and RR over 1,000 random combinations of trels, as a function of pool size.

# 7. Conclusions

The first edition of the Information Retrieval Education through Experimentation (EIREX) was run as an attempt to bring TREC-like evaluations to the IR undergraduate classroom. Doing so, we get students involved in the whole process of building a search engine and a test collection to evaluate it. Our goal is to introduce students in this kind of laboratory experiments in Computer Science, with a special focus on how to evaluate their systems and analyze the results. We have described how to adapt TREC's ad-hoc methodology to build such collections for an IR course. The first main difference is that the documents in the collection are gathered after selecting the topics, and not the other way around as usual. The second main difference is related to the pools of documents to judge: the systems developed by the students cannot contribute directly to the pools to prevent cheating, and the judging effort is limited because the students cannot be asked to judge as many documents as we would want. Due to this limitation, the pools are formed differently, with the help of freely available IR tools.

The question is whether such small-scale experiments are reliable or not, which is again an excellent question to investigate with the students, so they learn how to analyze them from a critical point of view to look into possible threats to validity [Voorhees, 2002][Urbano, 2011]. The main threats to validity in our case are the inconsistency and incompleteness of relevance judgments, so we measured the reliability of our methodology with typical meta-analysis techniques. We observed high agreement scores between students, and very high correlations between system rankings when using different sets of relevance judgments. In terms of incompleteness, we estimated that pools of size 100 and different depths are quite reliable and do not seem to affect the evaluation significantly. We conclude that the judgments made by students can be trusted, and that the pooling method proposed seems to work reasonably well for these small-scale evaluations.

In the future, we plan to keep building one new test collection each year. Given that the inconsistency of judgments does not significantly affect the outcome of the experiments, we will try to put our efforts into having more topics. In addition, we plan on having students propose the topic set themselves, with per-topic description of what is considered relevant. A related issue is trying to develop a topic set such that the similarity of documents between topics is larger and so finding the relevant ones is harder. We also plan on studying the use of better quality control techniques upon the relevance judgments made by students, looking at it as a typical crowdsourcing task.